# Probing the role of $Nd^{3+}$ ions in the weak multiferroic character of $NdMn_2O_5$ by optical spectroscopies

S. Mansouri[1,2*], S. Jandl[1], M. Balli[1], P. Fournier[1,3], B. Roberge[4], M. Orlita[4], I. A. Zobkalo[5], S.N. Barilo[6], S.V. Shiryaev[6]

[1]Institut Quantique et Regroupement québécois sur les matériaux de pointe, Département de Physique, Université de Sherbrooke, Sherbrooke, J1K 2R1 Canada
[2]Département de Chimie, Université de Sherbrooke, Sherbrooke, J1K 2R1 Canada
[3]Canadian Institute for advanced Research, Toronto, Ontario, M5G1Z8, Canada
[4]Grenoble High Magnetic Field Laboratory, 25, Avenue des Martyrs, Boîte Postale 166 F-38042 Grenoble, France
[5]Petersburg Nuclear Physics Institute, NRC Kurchatov Institute, Orlova Roshcha, Gatchina, St. Petersburg district 188300, Russia
[6]Scientific-Practical Materials Research Centre NAS of Belarus, 19 P. Brovki str., Minsk, 220072, Belarus

*Corresponding author: saber.mansouri@usherbrooke.ca

## Abstract

Raman and infrared spectroscopies are used as local probes to study the dynamics of the Nd-O bonds in the weakly multiferroic $NdMn_2O_5$ system. The temperature dependence of selected Raman excitations reveals the splitting of the Nd-O bonds in $NdMn_2O_5$. The $Nd^{3+}$ ion crystal field (CF) excitations in $NdMn_2O_5$ single crystals are studied by infrared transmission as a function of temperature, in the 1800-8000 cm$^{-1}$ range, and under an applied magnetic field up to 11 T. The frequencies of all $^4I_j$ crystal-field levels of $Nd^{3+}$ are determined. We find that the degeneracy of the ground-state Kramers doublet is lifted ($\Delta_0$ ~7.5 cm$^{-1}$) due to the $Nd^{3+}$-$Mn^{3+}$ interaction in the ferroelectric phase, below $T_C$ ~ 28 K. The $Nd^{3+}$ magnetic moment $m_{Nd}(T)$ and its contribution to the magnetic susceptibility and the specific heat are evaluated from $\Delta_0(T)$ indicating that the $Nd^{3+}$ ions are involved in the magnetic and the ferroelectric ordering observed below ~ 28 K. The Zeeman splitting of the excited crystal field levels of the $Nd^{3+}$ ions at low temperature is also analyzed.

# I. Introduction

Manganite multiferroic materials, $RMn_2O_5$ and $RMnO_3$ (R = Tb, …Tm), have recently attracted much attention due to their potential for technological applications [1-5]. For example, they offer the opportunity to realize a **M**agneto**e**lectric **R**andom **A**ccess **M**emory (MeRAM) [6] that would allow writing magnetic information by the application of small electric fields. They are also candidate materials for the liquefaction of hydrogen and helium gases using a giant rotating magnetocaloric effect at low magnetic fields as recently reported in the multifunctional materials $TbMn_2O_5$, $HoMn_2O_5$ and $TbMnO_3$ single crystals [3-5]. In order to implement these systems in functional devices or conceive new optimized multifunctional materials, one needs to elucidate the mechanisms behind their physical properties, in particular the challenging issue of the microscopic origin of the spin-induced ferroelectricity.

The microscopic mechanisms usually proposed to explain the spin-induced ferroelectricity in orthorhombic $RMnO_3$ are i) the antisymmetric Dzyaloshinskii-Moriya (DM) interaction between non-collinearly ordered spins, which breaks the inversion symmetry through the displacement of negatively charged ligands [7], or ii) by the spin current model [8]. In contrast, the ferroelectricity in $RMn_2O_5$ systems, induced or enhanced by a nearly collinear magnetic interaction, cannot be explained by the standard DM model. This failure has opened a debate on the microscopic origin of ferroelectricity in this family [1, 9-11]. The strong magnetoelectric coupling in the $RMn_2O_5$ series with polarization along their b axis can be as large as $P_{//b} > 3000$ $\mu Cm^{-2}$ for $GdMn_2O_5$, while, for the orthorhombic $RMnO_3$ systems, the largest polarization reaches only $P_{//c} \sim 600$ $\mu Cm^{-2}$ for $TbMnO_3$ along its c axis. For both families, this magnetoelectric coupling is also strongly dependent on the size of the rare-earth [2, 12-15]. The $RMn_2O_5$ compounds with large ionic radii (R= La and Pr) do not exhibit a detectable electric polarization and are considered as paraelectrics [14], while those with small ionic radii (R=Sm to Lu) display a finite electric polarization [13]. The intermediate size member of this family $NdMn_2O_5$ represents a particular case between the non-ferroelectric $PrMn_2O_5$ and the ferroelectric $SmMn_2O_5$. Recently, Chattopadhyay *et al.* [16-17] have revealed the weak ferroelectric character of this compound. The electric polarisation of $NdMn_2O_5$ (~2.4 $\mu Cm^{-2}$) is two orders of magnitude smaller than that of other multiferroic members of this series. Moreover, unlike the other multiferroic members of this family, its electrical polarization arises in an incommensurate magnetic state [13].

The $RMn_2O_5$ compounds crystallize in the orthorhombic structure [18-19] where edge-shared $Mn^{4+}O_6$ octahedra are connected along the *c*-axis and pairs of $Mn^{3+}O_5$ pyramids are linked to two $Mn^{4+}O_6$ chains. The rare earth ions are located in distorted $RO_8$ polyhedra. However, there is no consensus on their space group symmetry. Indeed, a slight deviation from the Pbam space group has been recently observed in some compounds of this family [20]. The magnetic frustration of these systems is imposed by the geometric configuration. The loop of five adjacent $Mn^{3+}$ and $Mn^{4+}$ ions leads to a complex magnetic configuration in the $RMn_2O_5$ systems [1, 21-22]. The $Mn^{3+}$ and $Mn^{4+}$ moments are coupled antiferromagnetically by the exchange parameters $J_4$ along the *a*-axis and $J_3$ along the *b*-axis [22]. The $Mn^{3+}$ moments, in two linked pyramids, couple antiferromagnetically via the exchange parameters $J_5$. As a result of these numerous exchange parameters and the magnetic frustration, $RMn_2O_5$ multiferroics with small R ions show multiple phase transitions as a function of temperature. A magnetic transition to an antiferromagnetic order with an incommensurate propagation vector is observed above ~ 40-45 K. At ~35-38 K, they show a commensurate magnetic order with a spontaneous electric polarization parallel to the b-axis. Close to ~20-25 K, a transition to a weakly ferroelectric state with an incommensurate magnetic order occurs. This last transition is followed by a rare-earth magnetic order at lower temperatures [1, 22].

$NdMn_2O_5$ undergoes also successive phase transitions driven by different intertwined magnetic and ferroelectric states [16-17]. Around $T_1$ ~ 31 K ($T_N$), an incommensurate magnetic order within a paraelectric phase (ICM1/PE) appears with two incommensurate magnetic propagation vectors. At $T_2$ ~ 26 ± 2 K ($T_C$), a ferroelectric-like state emerges while the incommensurate magnetic order persists (ICM1/FE). As temperature continues to decrease, it is followed by a lock-in type transition around $T_3$ ~ 15 K with a different incommensurate magnetic order within the same ferroelectric phase (ICM2/FE). Finally, at $T_4$ ~ 4 K, an additional magnetic transition is attributed to the ordering of $Nd^{3+}$ magnetic moments with respect to the moment of the manganese sublattice. Interestingly, the occurrence of the electric polarization only in the ICM phases, in contrast to the other $RMn_2O_5$ compounds, renders it unique and demands further microscopic investigation measurements to provide a deeper understanding on the origin of its FE properties and the role of $Nd^{3+}$ magnetic moments on the magnetoelectric transitions.

Raman scattering and infrared measurements have been used successfully for the study of the interplay of charge, spin, and lattice degrees of freedom in manganite multifunctional materials [23-27]. The study of the phonon and CF excitation evolutions as a function of temperature and/or magnetic field provides interesting information regarding the magnetic and the electronic properties as well as the surrounding ligands and the local inhomogeneities [28]. In particular, the $Nd^{3+}$, $Sm^{3+}$ and $Dy^{3+}$ Kramers ions, with odd numbers of electrons in their 4f shells, have levels with two-fold degeneracies that could be lifted in the presence of local and applied magnetic fields. In fact, the high sensitivity of the Kramers-ion crystal-field levels to both the local electric and magnetic fields, combined with the high resolution of the optical techniques, makes it a valuable probe for existing magnetic exchanges. To our knowledge, there is no information in the literature about the $Nd^{3+}$ crystal-field scheme for $NdMn_2O_5$.

In this paper, we present a highly detailed study of the $Nd^{3+}$ CF excitations in $NdMn_2O_5$ single crystals, as a function of temperature and under applied magnetic field (at T = 4.2 K). These experiments are complemented with Raman and magnetization measurements of $NdMn_2O_5$. The main objective of this paper is to probe the role of $Nd^{3+}$ ions in the weak multiferroic $NdMn_2O_5$ by studying its Kramers-doublet (KD) excitation responses as a sensitive probe of magnetic interactions and external magnetic field as well as the local disorder of the surrounding ligands.

## II. Experiment

$NdMn_2O_5$ single crystals used for these measurements were grown with the flux-melt method as described in Ref. [29]. These crystals show thin rectangular platelet shapes, with typical dimensions of 2 x 2 x 0.3 $mm^3$. The Raman spectra were measured using a Labram-800 Raman spectrometer equipped with an appropriate notch filter and a nitrogen-cooled CCD detector. The samples were mounted on the cold finger of a micro-helium Janis cryostat and the Raman spectra were recorded between 5 K and 300 K. The excitation laser line (632.8 nm) was focused through a 50x objective with intensity less than 0.8 mW to avoid local heating. The infrared transmission spectra were recorded in the 1500-8000 cm$^{-1}$ range with a Fourier transform interferometer BOMEM DA3.002 equipped with a quartz-halogen source, a $CaF_2$ beamsplitter and an InSb detector. In this measurement configuration, the incident light propagates parallel to the *c*-axis. For experiments in magnetic fields, the sample was placed in a cryostat equipped with a superconducting coil with the applied magnetic field along the *c*-axis and parallel to incident light propagation direction and was probed via light-pipe optics using a Bruker Vertex 80v infrared spectrometer. The sample was kept at 4.2 K in a low-pressure helium gas during the experiments. The magnetization measurements were carried out using a superconducting quantum interferometer device (SQUID) magnetometer from Quantum Design.

### III. Experimental results

Vibrational spectroscopy signatures reflect the space group symmetry and informs on the local structural properties of materials based on the characteristics of the excited phonons. Figure 1 shows the temperature dependence of the Raman spectra of $NdMn_2O_5$ in the x(y'z')x ≡ $k_i(e_ie_s)k_s$ configuration. In the notation $k_i(e_ie_s)k_s$, the letters k and e denote the propagation vectors and the polarizations of the incident (i) and scattered (s) light, respectively. The Raman spectra obtained in the x(zz)x and x(yz)x configurations are presented in the inset as a confirmation of crystal symmetry. Each set of Raman spectra is recorded under the same experimental conditions and on the same spot of the sample. The typical phonons associated with the orthorhombic $RMn_2O_5$ manganites are observed [24, 28, 30-32]. The assignment of the phonon symmetries refers to the previous studies in Refs.

31 and 32. For all detected phonons, the linewidths are close to 3-9 cm$^{-1}$ confirming the high crystalline quality of our samples. With decreasing temperature, between 300 K and 5 K, no additional modes appear indicating a structural phase stability of the samples. Figures 2(a), 2(b) and 2(c) show respectively the thermal evolution of the Raman frequencies, the normalized integrated intensities and the full widths at half maximum (FWHM) of the 214, 601, 622, 653, 681 cm$^{-1}$ modes for NdMn$_2$O$_5$. Their temperature dependence exhibits subtle anomalies around the characteristic temperatures, $T_C \sim 28$ K, $T^* \sim 70$ K and $T^S \sim 150$ K. Interestingly, the Raman frequency of the $\sim 622$ cm$^{-1}$ mode follows the anharmonic model between 300 K and 160 K but then decreases below $T^S \sim 150$ K. The Raman intensities of the modes at 214, 601, 622, 681 cm$^{-1}$ decrease significantly below $T^S \sim 150$ K. Similar thermal behavior has also been observed for the Raman modes at $\sim 220$ cm$^{-1}$, $\sim 327$ cm$^{-1}$ and $\sim 700$ cm$^{-1}$ as we have recently reported in the TbMn$_2$O$_5$, HoMn$_2$O$_5$ and YMn$_2$O$_5$ compounds [28]. We have attributed this characteristic temperature at $T^S \sim 150$ K to a local disorder effect induced by the splitting of R-O bonds, into short and long bonds. This thermal disorder is reduced at $T^* \sim 70$ K when the number of short bonds matches up the number of the long bonds in the unit cell [28].

Figure 3(a) presents the DC magnetic susceptibility M/H of the NdMn$_2$O$_5$ single crystal, measured parallel to the *a*- and *c*-axis under a magnetic field of 500 Oe. The magnetic susceptibility increases with decreasing temperature up to 4.5 K and then decreases. This anomaly is usually attributed to a long-range antiferromagnetic order of the Nd$^{3+}$ moments, similar to the ones found in other compositions of the RMn$_2$O$_5$ series [1, 28, 33]. The inset shows their corresponding inverse susceptibility ($\chi^{-1}$). Well above $T_N$, the inverse susceptibility follows well the Curie-Weiss behavior $1/\chi = T/C - \theta_{CW}/C$ (dashed lines), yielding an effective magnetic moment value $\mu_{ob} \approx 6.45$ $\mu_B$ that is close to the expected value $\mu_{eff} = \sqrt{\mu_{eff}^2(Nd^{3+}) + \mu_{eff}^2(Mn^{3+}) + \mu_{eff}^2(Mn^{4+})} = 6.22 \mu_B$. The important negative value of $\theta_{CW} \sim -161$ K (along the $c$-axis) suggests dominant antiferromagnetic interactions and the large ratio of $|\theta_{CW}|/T_N \sim 5$ indicates a strong spin frustration and an strong magnetic correlation well above $T_N$. Also, $\chi^{-1}$ starts to deviate significantly from the Curie-Weiss behaviour at temperatures below $T^S$ and $\chi^{-1}(H//c) - \chi^{-1}(CW)$ presents a relative maximum at $T^* \sim 70$ K. This implies that the magnetic properties of NdMn$_2$O$_5$ are sensitive to the dynamics of the Nd-O bonds. Similar behaviours are also observed in TbMn$_2$O$_5$ and BiMn$_2$O$_5$ as a signature of the onset of bond length disorder [28, 30]. The first derivatives of $\chi(T)$, measured with H = 500 Oe parallel to the $c$-axis and the $a$-axis, are shown in Figs. 3(b) and 3(c) respectively. Anomalies at 42 ±1 K and 27 ± 2 K are observed. The anomaly around 42 K vanishes at a magnetic field of 5000 Oe as a likely precursor of the long-range Mn magnetic order at $T_1 \sim 31$K. The anomaly at 27 ± 2 K coincides with the appearance of a weak spontaneous ferroelectric polarization reported in previous studies [16-17]. The isothermal magnetization versus applied magnetic field data along the $a$- (at 2 K and 8 K) and $c$-axis (at 2 K) are shown in Fig. 3(d). The inset shows the derivative of M (dM/dH) obtained along the $c$-axis. In small applied magnetic fields, the magnetization increases linearly with H, but in the region of ~ 2.5-3 Tesla, there is a kink in the M (H) dependence (more pronounced for **H // a**). This behavior is not observed for T ≥ 6 K indicating that the low-temperature magnetic structure undergoes a magnetic-field-induced transition around 2.5-3.0 Tesla [18]. For H ≥ 3 Tesla, the magnetization continues to increase and no tendency toward saturation is observed up to 7 Tesla.

In orthorhombic NdMn$_2$O$_5$, the environment and the interactions within the 4f$^3$ electron configuration split the Nd$^{3+}$ free-ion electronic levels. The level multiplets have total moments J = 9/2, 11/2, 13/2 and 15/2 respectively with their corresponding 5, 6, 7, and 8 Kramers doublets for the $C_s^{xy}$ site symmetry. Each $^4$I$_J$ multiplet corresponds to (2J+1)/2 sublevels or Kramers doublets. Figure 4(a) presents the transmission spectrum of NdMn$_2$O$_5$ at 4.5 K for **E** parallel to the *xy*-plane. Three groups of sharp absorption lines due to f-f transitions of Nd$^{3+}$, centered at 2200 cm$^{-1}$, 4000 cm$^{-1}$ and 6000 cm$^{-1}$ are detected. As theoretically predicted, these absorption bands are the I$_{9/2}$ → I$_{11/2}$, I$_{9/2}$ → I$_{13/2}$ and I$_{9/2}$ → I$_{15/2}$ CF transitions. In Fig. 4(b), the temperature evolution of the I$_{9/2}$ → I$_{13/2}$ CF transitions is presented. Above 28 K, the six I$_{11/2}$, the seven I$_{13/2}$ and the eight I$_{15/2}$ expected CF levels are observed in the 1950-2300 cm$^{-1}$, 3900-4275 cm$^{-1}$ and 5800-6375 cm$^{-1}$ range, respectively. Table 1 lists the CF energy levels of the 26 Kramers doublets of Nd$^{3+}$ in NdMn$_2$O$_5$ at 30 K. The I$_{9/2}$ ground state excited levels are determined when they are thermally populated, at high temperatures, above 100 K. Their energy values are deduced from the comparison between the thermally excited levels of the $^4$I$_J$ multiplets (J = 11/2, 13/2 and 15/2). A complete identification of the CF level frequencies is necessary for the success of the numerical search for reliable CF parameters that depends strongly on the initial estimate.

With decreasing temperature, the excitation widths decrease, with values reaching typically 20 cm$^{-1}$ at 30 K. Below 30 K, the degeneracy of the KD ground state is lifted due to the Nd-Mn exchange interaction following the antiferromagnetic ordering of the Mn moments and exactly at the occurrence of the ferroelectric phase transition at T=28 K. Indeed, all the CF levels of the $^4$I$_{13/2}$ multiplets split at the same temperature of T = 28 K with almost the same magnitude. Typically, at 20 K the doublet frequencies for the $I_{13/2}$ multiplets are: 3917-3923, 3973-3979, ~4008-4014, 4069-4075, 4114-4120, 4162-4168, 4242-4248 cm$^{-1}$. For the ground-state KD, as temperature is lowered from 28 to 4.5 K, the average doublet splitting increases from ~5.5 cm$^{-1}$ at 28 K to ~7.5 cm$^{-1}$ at 4.5 K.

The evolution of the KD splittings of the $I_{11/2}$, $I_{13/2}$ and $I_{15/2}$ multiplet levels as a function of magnetic field (**B** || c) at T= 4.2 K are shown in Figs. 5(a), 5(b) and 5(c) respectively. Additional KD splittings due to the Zeeman effect are observed. At 4.2 K, the high-energy Kramers-doublets component of the ground state is very weakly populated and does not contribute to the CF excitations. Hence the detected splittings are directly due to the KD excited multiplets. For the saturated bands, around 2040 cm$^{-1}$ and 4012 cm$^{-1}$, the two components of the KD excitation can be easily distinguished at higher magnetic fields when they are largely separated. For some CF excitations, such as those at 3978 cm$^{-1}$, 4077 cm$^{-1}$, 4123 cm$^{-1}$ and 4172 cm$^{-1}$, the magnitude of the KD splitting becomes more pronounced above 3 Tesla.

## IV. Discussion

A zoom of the temperature dependence of the absorption bands (~ 3920 cm$^{-1}$) for the f-f transition of Nd$^{3+}$ in NdMn$_2$O$_5$ is presented in Fig. 6(a). The temperature dependence of the frequency and the full width at half maximum (FWHM) of the 3920 cm$^{-1}$ CF excitation between 275 K and 30 K are presented in Figs 6(b) and 6(c) respectively. The frequency of the 3920 cm$^{-1}$ CF excitation hardens between 300 K and 200 K and becomes nearly constant between 200 K and 100 K and hardens again below 100 K. Interestingly, the FWHM of the 3920 cm$^{-1}$ CF excitation decreases progressively between 300 K and 100

K and increases considerably between 100 K and 30 K. Such behavior is also observed for the FWHM of the $Nd^{3+}$ CF excitations at 3975 cm$^{-1}$, 4074 cm$^{-1}$, 4168 cm$^{-1}$ and 4246 cm$^{-1}$. Broadenings of the CF excitations are similar to what we have recently reported for the $Tb^{3+}$ and $Ho^{3+}$ CF excitations in $TbMn_2O_5$ and $HoMn_2O_5$ [28]. They confirm a universal behaviour of the $RMn_2O_5$ lattice dynamic properties. These overall findings are attributed to a local disorder effect induced by the splitting of R-O bonds that also explains the decrease of the Raman intensities and the frequency shifts of some $NdMn_2O_5$ phonons below 150 K. The relationship between the thermal dynamics of the R-O bonds and the ferroelectric properties in the $RMn_2O_5$ systems at high temperatures [20, 34-35] needs to be further investigated. We presume that the thermal disorder effect induced by the splitting of R-O bonds involves a small displacement of the $R^{3+}$ ions from its centrosymmetric site in Pbam that complicates the determination of the actual space group symmetry of $RMn_2O_5$ systems at high temperature as reported in Ref. 20.

Below 28 K, the band at 3921 cm$^{-1}$ is resolved in bands at 3923.3 cm$^{-1}$ and 3917.7 cm$^{-1}$. As the temperature is lowered, below 20 K, these two bands split in two other close bands at 3916.2-3918.7 cm$^{-1}$ and 3922.3-3924.8 cm$^{-1}$ as a result of the splitting of the two Kramers doublets: the $Nd^{3+}$ KD ground state and the excited KD level at 3921 cm$^{-1}$ as illustrated in Fig. 6(d) where the four arrows indicate the possible transitions. The splitting magnitude of the KD excited state ($\Delta'$) is smaller than that of the KD ground state ($\Delta_0$). Such detailed scheme is not observed for other excited levels due to their relatively broad widths that gives an averaged picture. The change of spectral weight, between the excitations at 3923.3 cm$^{-1}$ and 3917.7 cm$^{-1}$, is due to the depletion of the ground state excited level KD with decreasing temperature. As expected, such change of spectral weight is not observed between the components of the excited KD.

Figure 7(a) shows the absorbance maps of $Nd^{3+}$ $I_{9/2} \rightarrow I_{13/2}$ CF transitions as a function of temperature, between 45 and 4.5 K. The absorbance spectra (A) are calculated from the transmission spectra (T) and the incident intensity ($I_0$) using the equation $A = -\log_{10}(T) = -\log_{10}\left(\dfrac{I}{I_0}\right)$. The empty white squares indicate the variation of the $^4I_{13/2}$ CF excitation frequencies as a function of temperature. The vertical dashed lines indicate the temperature of the different phase transitions. In the ferroelectric phase, below 28 K, the different lines are equally split compared to their positions at 30 K. In addition, all the observed lines behave the same way with decreasing temperature. The temperature dependence of the splitting shift is almost the same for all the CF excitations. Also, the spectral weight of each excitation components is shifted at higher frequency with decreasing temperature. All these observations indicate clearly that the main contribution to the line splitting arises from the lifted degeneracy of the KD ground state. The magnitude of the ground-state splitting at 4.5 K ($\Delta_0 \sim 7.5$ cm$^{-1}$) is less important than that observed in $NdMnO_3$ ($\Delta_0 \sim 14$ cm$^{-1}$) [36]. This could be related to the incommensurate magnetic structure of the Mn spins in $NdMn_2O_5$ which creates weak effective magnetic fields at different Nd sites compared to the canted ferromagnetic state in $NdMnO_3$.

The variation of $\Delta_0(T)$ as a function of temperature is presented in Figure 8(a). The $Nd^{3+}$ ground-state Kramers doublet splits below $T_C \sim 28$ K as the crystal gets into the ferroelectric phase. In the following, we use the $\Delta_0(T)$ dependence, obtained as described above, to estimate the Neodymium magnetic moment $m_{Nd}(T)$ and the Neodymium contribution to the specific heat and the magnetic susceptibility. We will compare these extracted properties with direct experimental data. The temperature dependence of the magnetization arising only from the $Nd^{3+}$ magnetic moments, $m_{Nd}(T)$, can be calculated within the framework of the ground-state doublet model, according to the following equation:

$$m_{Nd}(T) / m_{Nd}(0) = \tanh(-\Delta_0(T)/2k_B T) \quad (1)$$

where $\Delta_0(T)$ is the energy splitting between the two components of the KD ground state, $k_B$ is the Boltzmann constant and $m_{Nd}(0)$ is the zero-temperature magnetization. The variation of $m_{Nd}(T) / m_{Nd}(0)$ is shown in Fig. 8(b). Here we find that $m_{Nd}$ is not zero below $T_C = 28$ K. It implies that the Nd ions are involved in the magnetic and the ferroelectric ordering observed below 28 K, unlike the scenario recently proposed by Chattopadhyay *et al.* [16], and consistently with what has been observed in $HoMn_2O_5$ and $TbMn_2O_5$ where the Ho and the Tb spins are partially ordered even at 26 K and 27 K, respectively [21].

To additionally checking whether the phase transitions observed in the magnetization measurements and the heat capacity measurements are connected to the $Nd^{3+}$ ions or not, we have evaluated the Neodymium contributions to the magnetic susceptibility $\chi_{Nd}(T)$ and specific heat $C_{Nd}(T)$, according to equations (2) and (3) respectively [37-39]:

$$\chi_{Nd}^a(T) = N_A \frac{m_{Nd}^a(0)^2}{k_B T} \frac{1}{\cosh^2(\Delta_0(T)/2k_B T)} \quad (2)$$

$$C_{Nd}(T) = R \left(\frac{\Delta_0(T)}{2k_B T}\right)^2 \frac{1}{\cosh^2(\Delta_0(T)/2k_B T)} \quad (3)$$

where $N_A$ is the Avogadro number and $R$ is the ideal gas constant. $m_{Nd}^a(0)$ is the $a$-component of the magnetization arising from Neodymium magnetic moments. Here, it is estimated to be equal to 1.2 $\mu_B$ [40]. The resulting curves of $\chi_{Nd}(T)$ and $C_{Nd}(T)$, are shown as solid blue connected circles in Figs. 8(c) and 8(d) respectively, where they are compared with the $\chi(T)$ magnetic susceptibility (our measurements) and the $C(T)/T$ specific heat of $NdMn_2O_5$ (taken from Ref. 17). Contributions of the Neodymium moments retrace well the low-temperature profiles of $\chi(T)$ and $C(T)/T$ of $NdMn_2O_5$. Interestingly, the subtraction of the calculated $\chi_{Nd}(T)$ (obtained from optical experiment) from the direct measurement of $\chi(T)$ of $NdMn_2O_5$ shown in the inset of Fig 8(c) as $\chi_{Mn}(T)$ reveals the signature of the antiferromagnetic order of the Mn ions below 30 K. $\chi_{Mn}(T)$, has a similar temperature profile and the same order of magnitude as the measured $\chi(T)$ of $YMn_2O_5$ and $BiMn_2O_5$ along the $a$-axis at low temperature [9, 41] for which Y and Bi do not carry a magnetic moment. Also, the Schottky-type Neodymium contribution in $C_{Nd}(T)/T$ explains well its increase at low temperature and the occurrence of a peak near 4.5 K corresponding to the $Nd^{3+}$ magnetic ordering [17]. A similar Schottky-type rare-earth anomaly could also explain the increase of $C(T)/T$ for $GdMn_2O_5$ and $DyMn_2O_5$ [42].

Fig 7(b) presents the absorbance map of $Nd^{3+}$ $I_{9/2} \rightarrow I_{13/2}$ CF transitions as a function of magnetic field at 4.2 K. As mentioned above, the Zeeman splitting of the CF excitations at 3978 cm$^{-1}$, 4077 cm$^{-1}$, 4123 cm$^{-1}$ and 4172 cm$^{-1}$ becomes more pronounced above 3 Tesla, exactly in the ICM2/FE phase. Coinciding with the transition observed in the magnetization measurements in Fig 3(d), this variation could be related to a magnetic-field-induced transition CM+ICM2/FE $\rightarrow$ ICM2/FE [18, 43] as a consequence of a renormalization of the KD g tensors. This renormalization increases with the level energy and could be helpful for the CF parameters modeling.

## Conclusion

In this paper, we have mainly presented a high-resolution spectroscopic investigation of $NdMn_2O_5$ single crystals in a wide range of wave numbers, temperature and applied magnetic field. We have reported the first CF study of $Nd^{3+}$ ion $4f^3$ electron excited states in $NdMn_2O_5$. All the CF energy levels of the 26 Kramers doublets of $Nd^{3+}$ have been determined which offers empirical data for future reliable CF parameters estimations. Interestingly, we have found that the ground-state Kramers doublet splits into two sublevels below 28 K, in the ICM/FE phase. This splitting $\Delta_0(T)$ increases below 20 K and reaches a maximum at 4.5 K. Using $\Delta_0(T)$, we have calculated the Neodymium magnetic moments $m_{Nd}(T)$ and the related contributions to the magnetic susceptibility and the specific heat. Our results reveal the antiferromagnetic order of Mn ions and give an adequate estimate of the Schottky-type contribution in specific heat. Analyzing all these experimental results, we infer that the Nd ions are involved in the weak ferroelectric properties of $NdMn_2O_5$, observed below ~ 28 K, via the Nd-Mn interaction consistently with what has been observed in $HoMn_2O_5$ and $TbMn_2O_5$ where the Ho and Tb moments are partially ordered even at 27 K. In addition to the detection of the ground-state Kramers doublet splitting at 28 K, evolution of the infrared active excitations $^4I_{9/2} \rightarrow {}^4I_{11/2}$, $^4I_{13/2}$ and $^4I_{15/2}$ under applied magnetic field retraces the magnetic-field-induced transition CM+ICM2/FE $\rightarrow$ ICM2/FE around 2.5-3 Tesla.

## Acknowledgment

We acknowledge the technical support of M. Castonguay, S. Pelletier, J. Rousseau and B. Rivard. SM, SJ, MB and PF acknowledge the support from the National Science and Engineering Research Council of Canada under grants RGPIN-2013-238476 (PF) and RGPIN-2012-350 (SJ), the Fonds Québécois de la Recherche sur la Nature et les Technologies and Canada Foundation for Innovation. This research was undertaken thanks in part to funding from the Canada First Research Excellence Fund. IAZ is grateful to the Russian Foundation for Basic Research for the financial support (grant No. 16-02-00545-a).

**Table captions**

Table I: The experimental CF excitation energies (in cm$^{-1}$) at 30 K of the Kramers doublets of Nd$^{3+}$ in NdMn$_2$O$_5$

**Table I**

| CF levels | $^4I_{9/2}$ | $^4I_{11/2}$ | $^4I_{13/2}$ | $^4I_{15/2}$ |
|---|---|---|---|---|
| Energy (cm$^{-1}$) | 0 | 1978 | 3922 | 5834 |
| | 142* | 2035 | 3975 | 5900 |
| | 238* | 2043 | 4011 | 6104 |
| | 291* | 2055 | 4074 | 6174 |
| | 329* | 2142 | 4120 | 6196 |
| | | 2257 | 4168 | 6262 |
| | | | 4246 | 6289 |
| | | | | 6344 |

* The energy of these CF excitations is determined at high temperature when they are thermally populated.

**Figures captions**

Figure 1: The temperature dependence of the Raman spectra of NdMn$_2$O$_5$ in the x(y'z')x. Inset: the Raman spectra of NdMn$_2$O$_5$ at 5 K measured in the x(zz)x and x(yz)x configurations.

Figure 2: Temperature dependence of the Raman frequency (a), the normalized integrated intensity (b) and the full width at half maximum (FWHM) (c) of the 214, 601, 622, 653, 681 cm$^{-1}$ modes in NdMn$_2$O$_5$. The dotted lines correspond to the expected anharmonic behavior. The characteristic temperatures T$_C$, T$_N$, T$^*$ and T$^S$ are defined in the text.

Figure 3: Temperature dependence of the magnetic susceptibility M/H of a NdMn$_2$O$_5$ single crystal measured in zero-field-cooled (ZFC) and field-cooled (FC) conditions with an applied magnetic field of 500 Oe along the *a* and the *c*-axis. Inset: The Curie-Weiss fitting of their inverse susceptibility as a function of temperature. (b) and (c) present the derivative of χ along the *c*- and the *a*-axis respectively. (d) Isothermal M(H) curves measured along the *a*-axis (for 2 K and 8 K) and the *c*-axis (for 2 K). Inset: The derivative of M(H) curve measured along the *c*-axis.

Figure 4: (a) Transmission spectra of Nd$^{3+}$ $^4$I$_{9/2}$ → $^4$I$_{11/2}$, $^4$I$_{3/2}$ and $^4$I$_{15/2}$ CF transitions in NdMn$_2$O$_5$ at 4.5 K. (b) The temperature dependence of the Nd$^{3+}$ I$_{9/2}$ → I$_{13/2}$ CF transitions in NdMn$_2$O$_5$. * indicates a saturated excitation.

Figure 5: Magnetic field dependence of the Nd$^{3+}$ I$_{9/2}$ → I$_{11/2}$ (a), I$_{9/2}$ → I$_{13/2}$ (b) and I$_{9/2}$ → I$_{15/2}$ (b) CF transitions in NdMn$_2$O$_5$ at 4.2 K. * indicate saturated excitations.

Figure 6: (a) The temperature dependence of the absorption bands (~ 3920 cm$^{-1}$) for the f-f transitions of Nd$^{3+}$ in NdMn$_2$O$_5$. (b) and (c) present the temperature dependence of the frequency and the full-width at half-maximum (FWHM) of the 3920 cm$^{-1}$ CF excitation between 275 and 30 K respectively. (d) presents a schematic of the splitting of the Nd$^{3+}$

ground-state and excited level of the Kramers doublet. Arrows show the observed absorption lines.

Figure 7: Absorbance maps of $Nd^{3+}$ $I_{9/2} \rightarrow I_{13/2}$ CF transitions as a function of temperature (a), between 45 K and 4.5 K, and magnetic field (b), where the empty white squares indicate the CF frequency levels of the $^4I_{13/2}$ multiplet. * indicates a saturated excitation. The notations PM, PE, ICM, CM and FE are used to represent paramagnetic, paraelectric, incommensurate magnetic, commensurate magnetic and ferroelectric states.

Figure 8: (a) The variation of the splitting $\Delta_0(T)$ as a function of temperature. (b) The temperature dependence of $m_{Nd}(T) / m_{Nd}(0)$ calculated from $\Delta_0(T)$ using Equation (1). (c) and (d) represent the Neodymium contribution to the magnetic susceptibility and the specific heat of $NdMn_2O_5$, calculated according to Equations (2) and (3) (blue filled circles) and compared with direct measurements (red filled circles) of magnetic susceptibility (H//a) and the specific heat (taken from the Ref. 17). Inset: difference in susceptibility $\chi_{NdMn2O5}(T) - \chi_{Nd}(T)$ corresponding to the Mn contribution to the magnetic susceptibility.

Figure 1

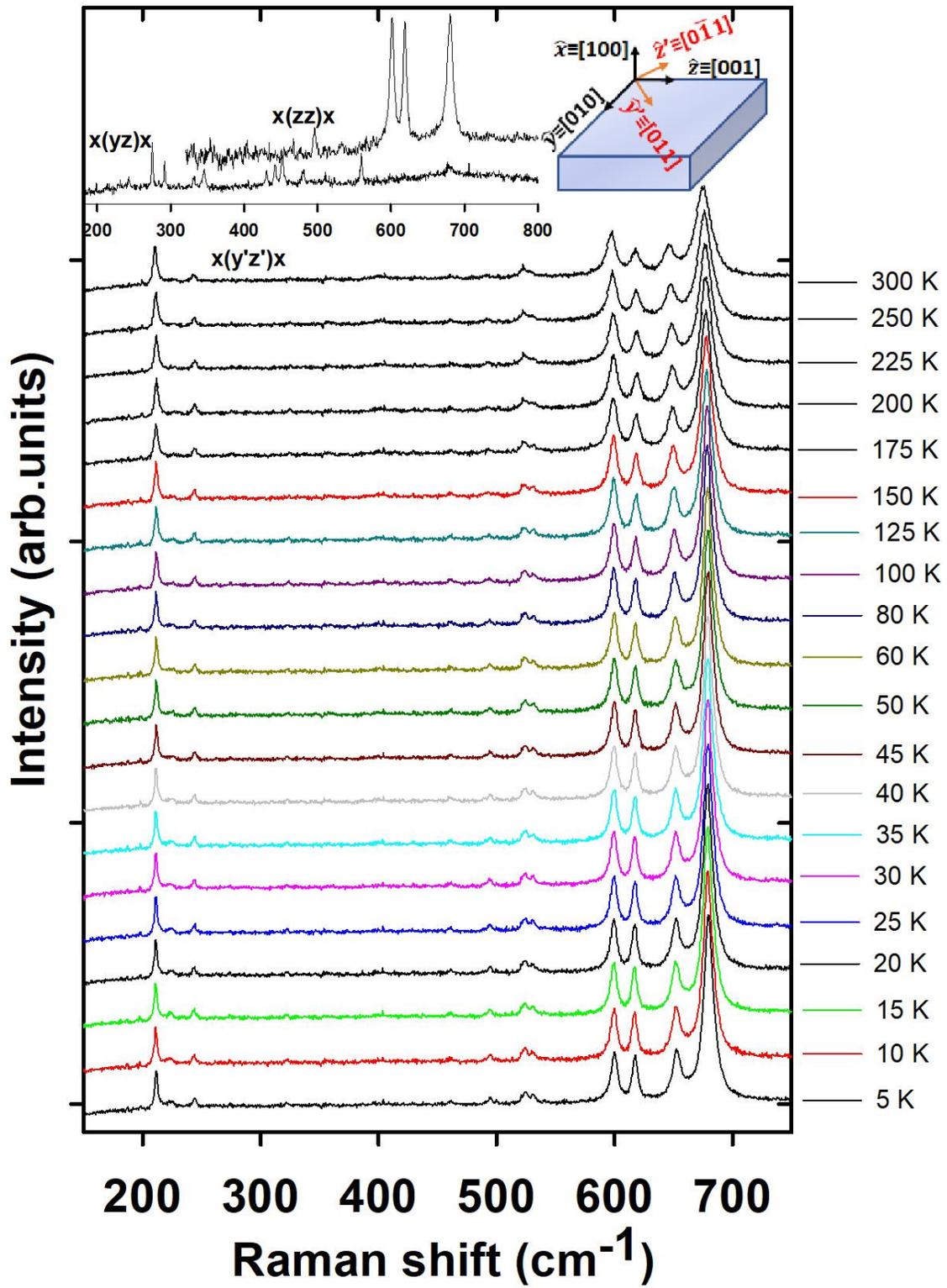

Figure 2

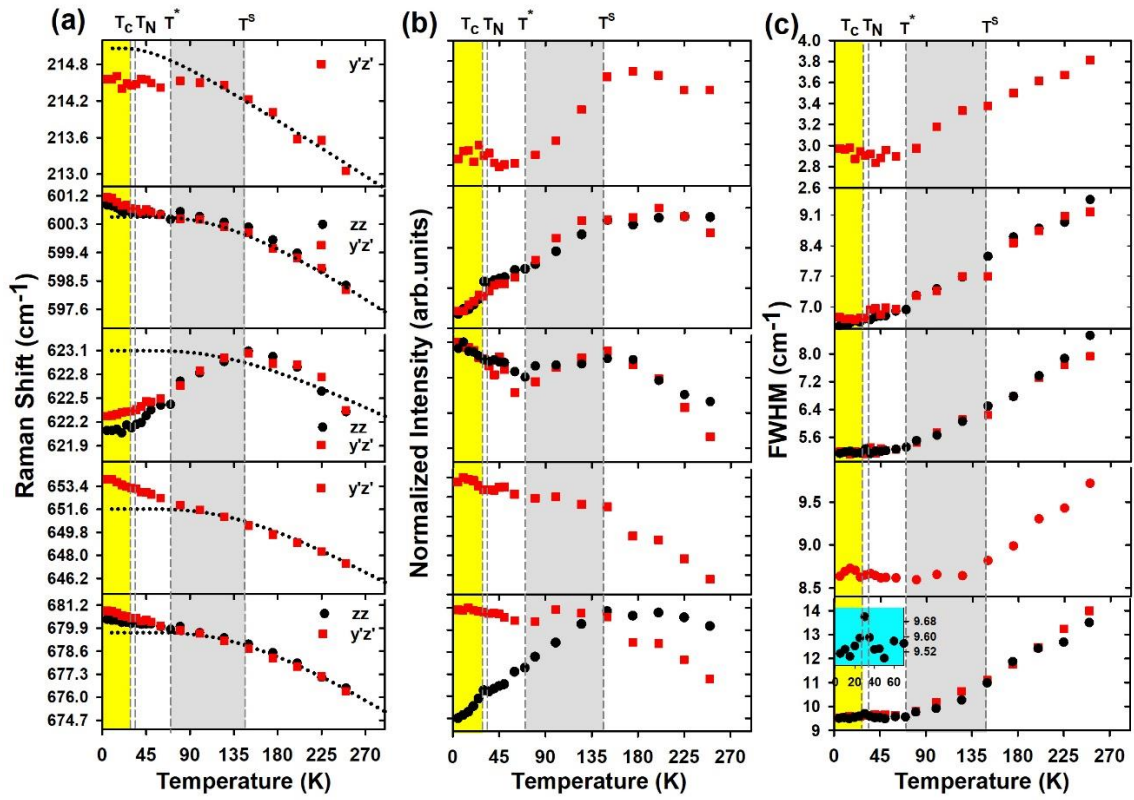

Figure 3

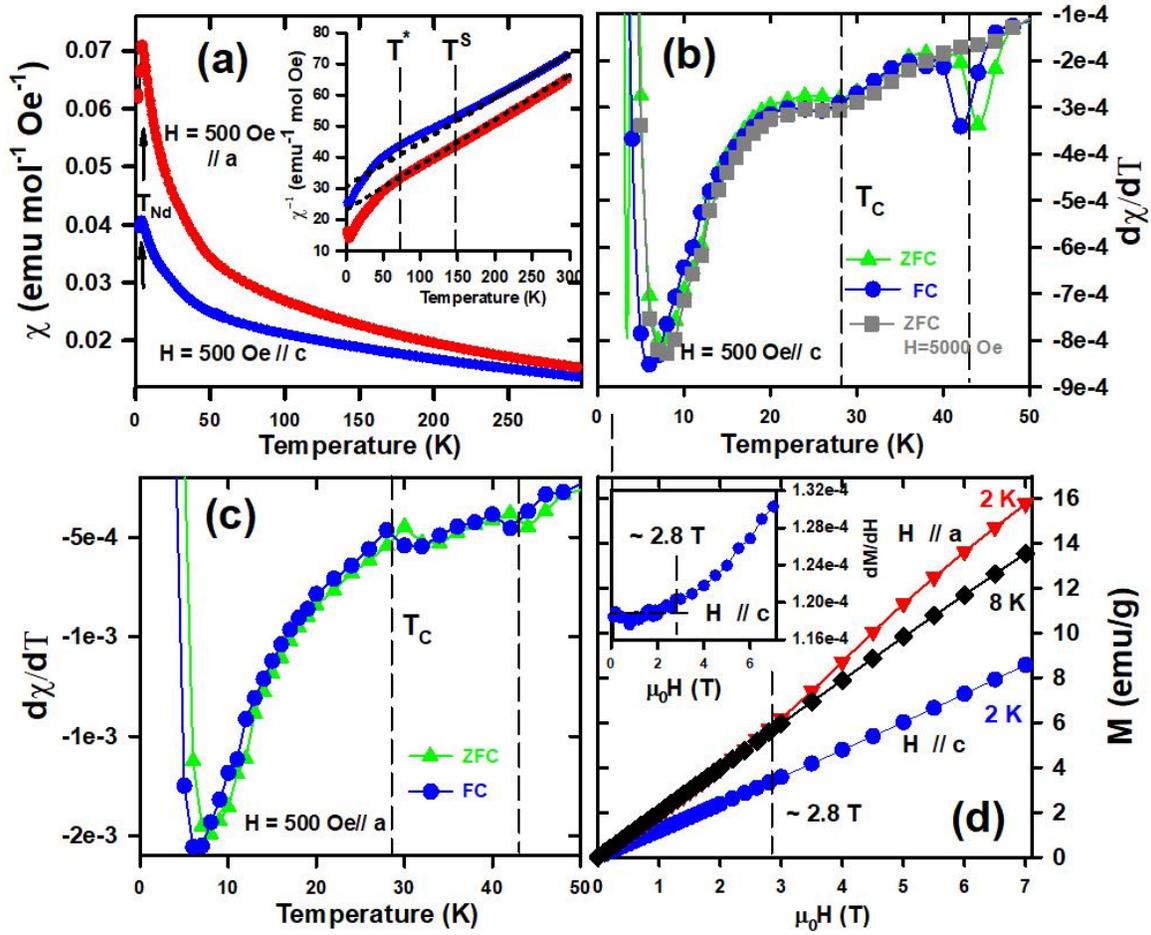

Figure 4

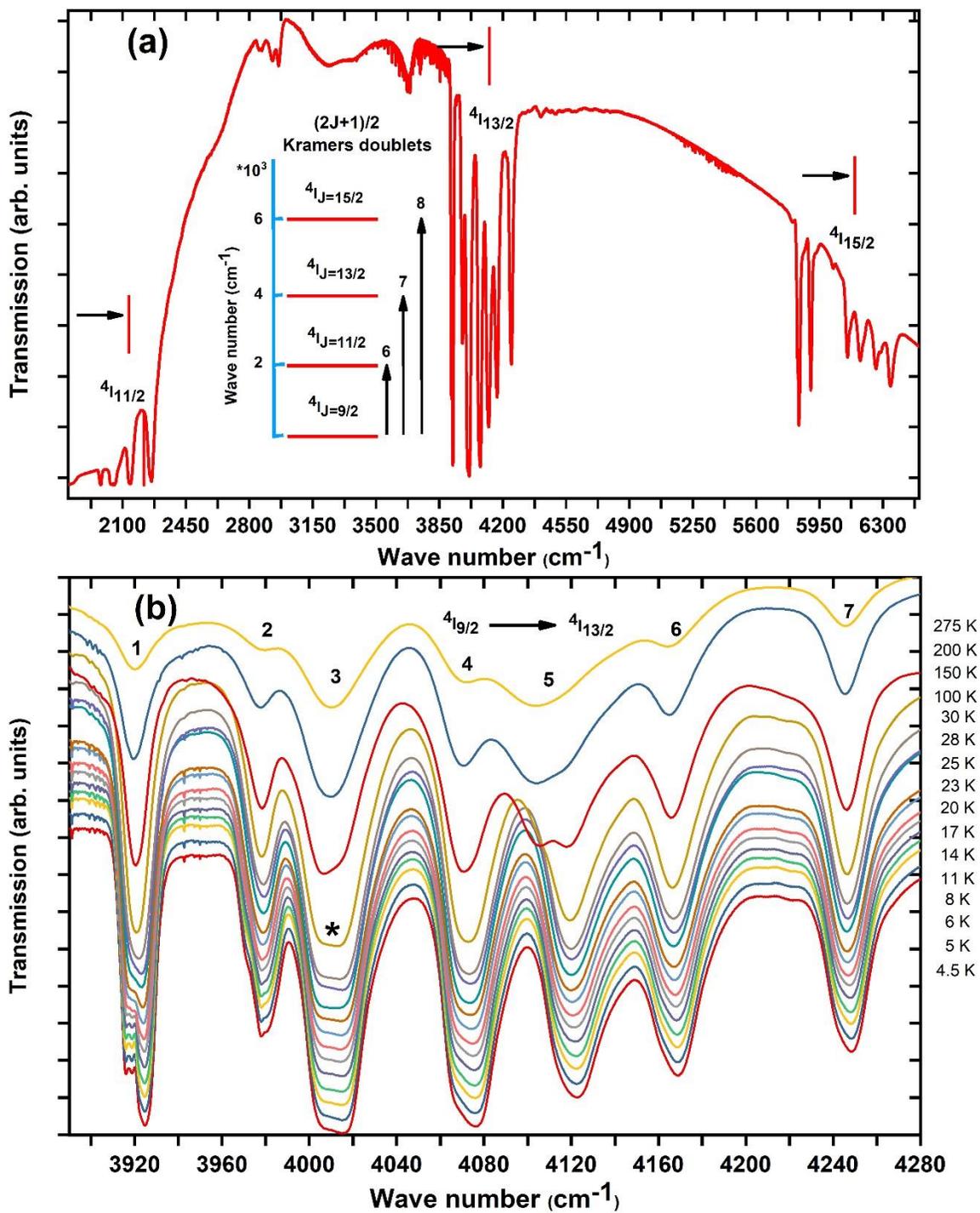

Figure 5

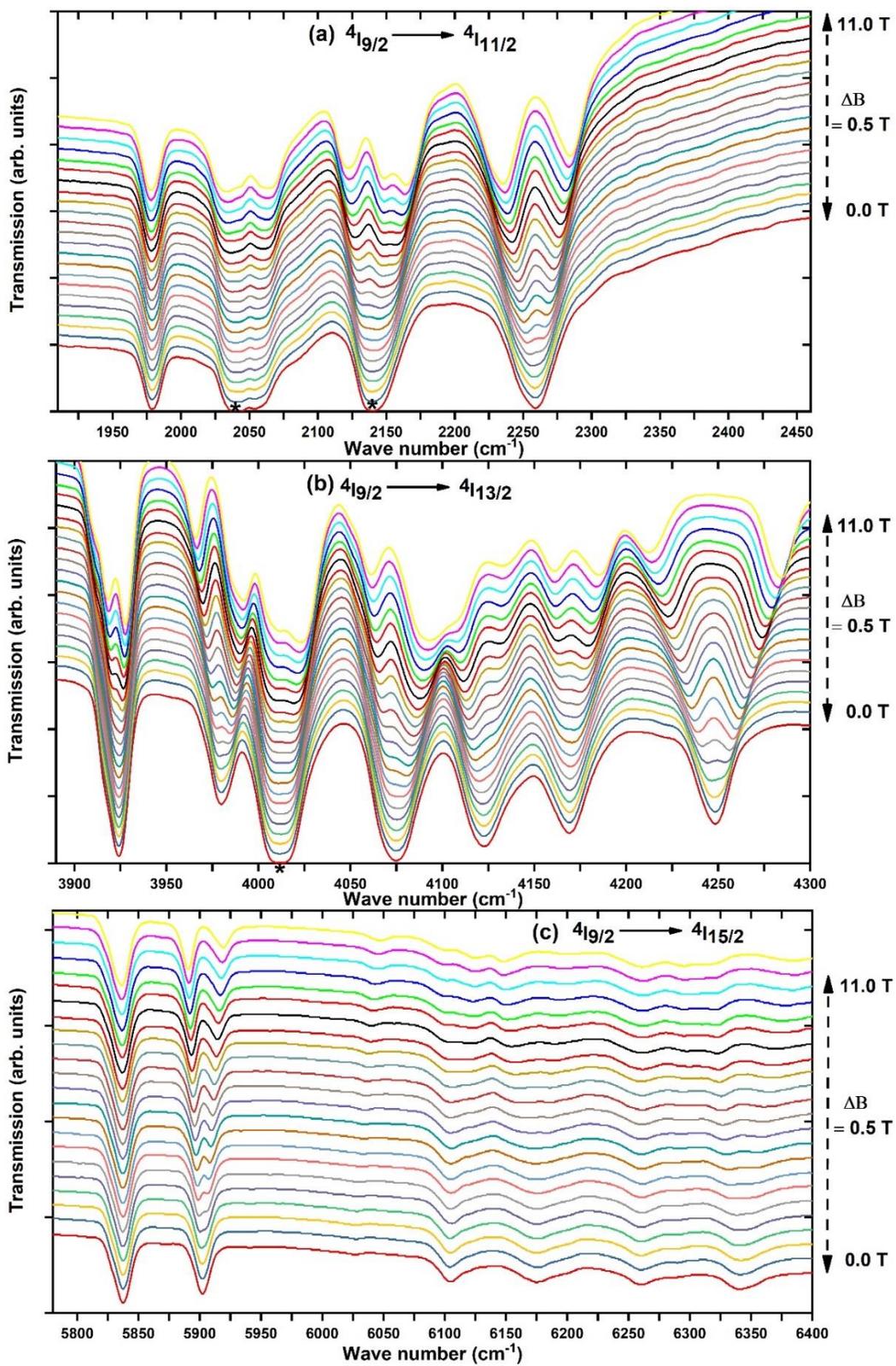

Figure 6

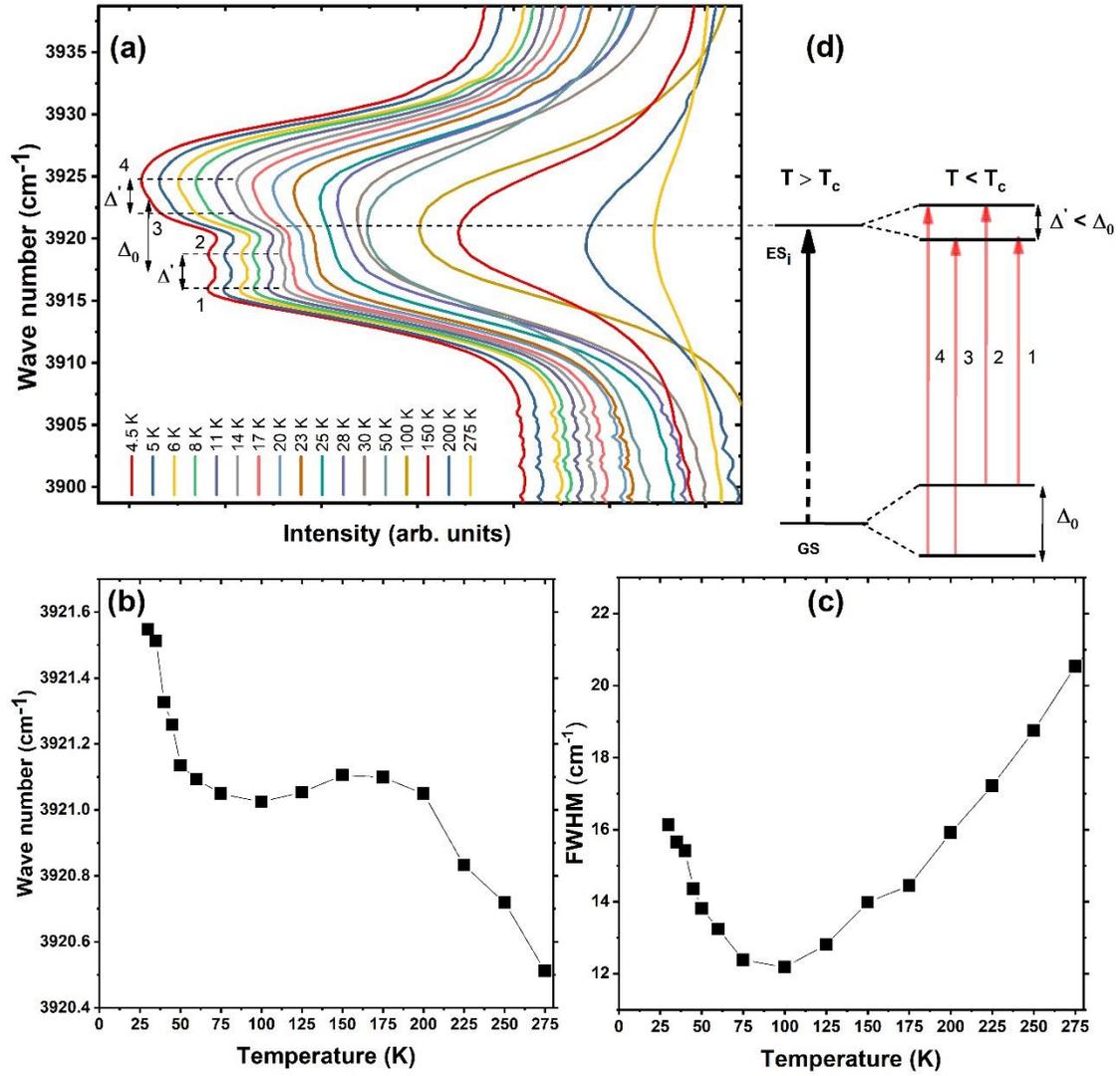

Figure 7

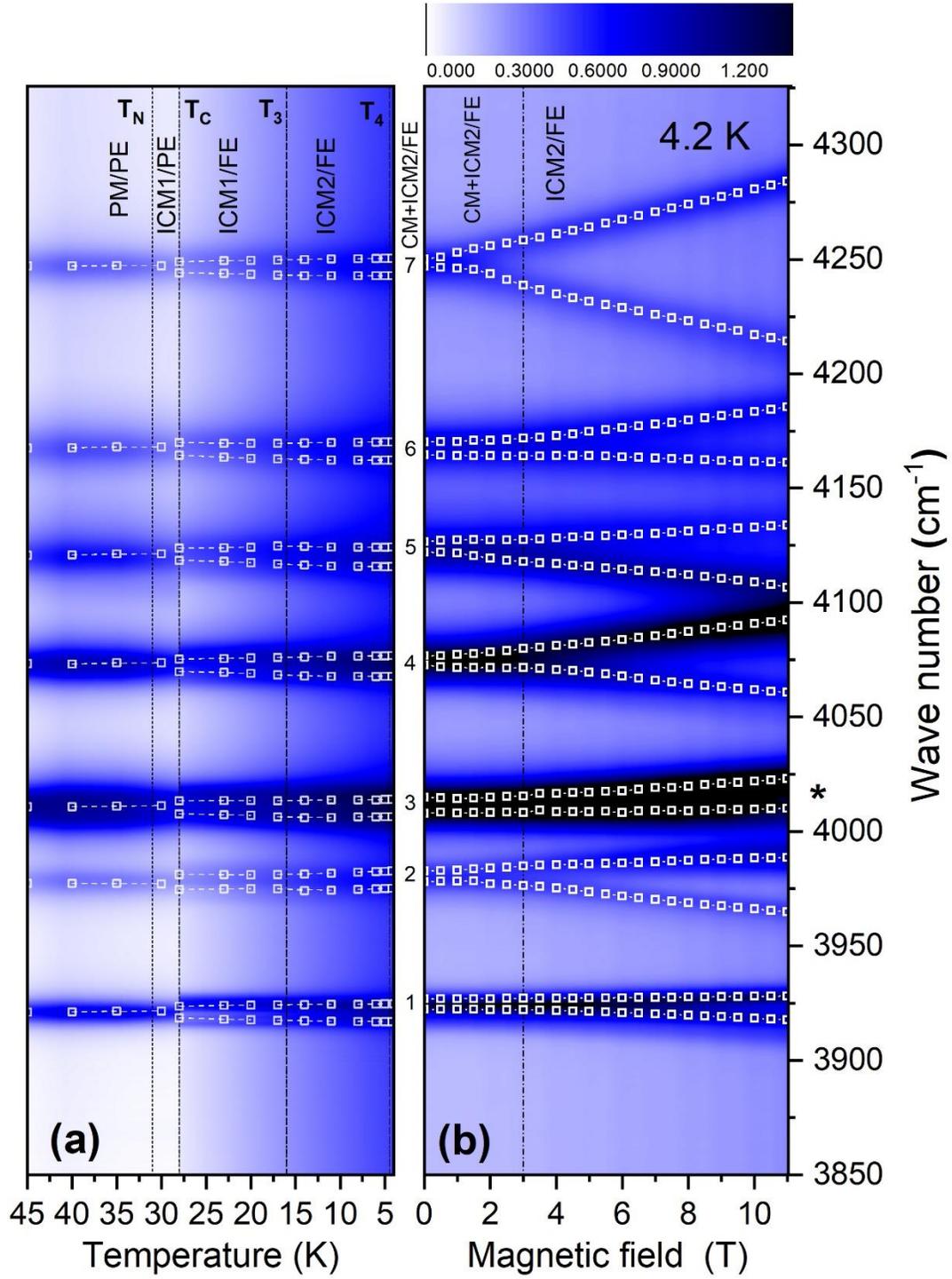

Figure 8

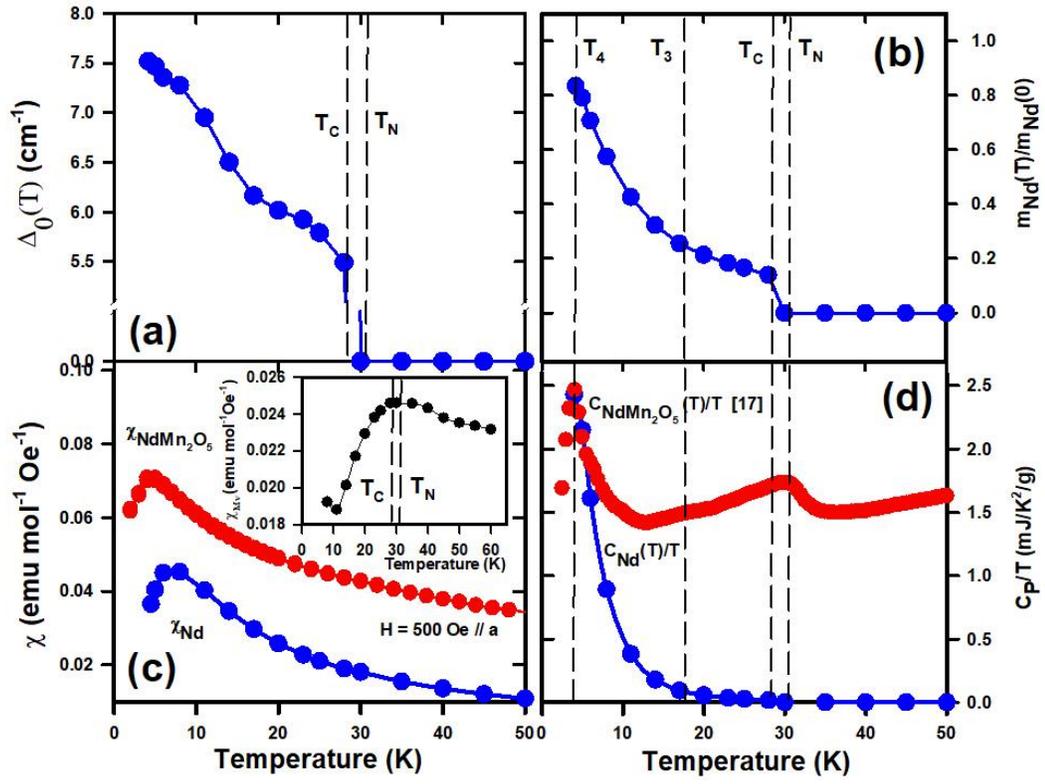